# Reversal of the extraordinary Hall effect polarity in thin Co/Pd multilayers.

D. Rosenblatt, M. Karpovski and A. Gerber

Raymond and Beverly Sackler Faculty of Exact Sciences

School of Physics and Astronomy,

Tel Aviv University

Ramat Aviv, 69978 Tel Aviv

Israel

Thin Co/Pd multilayers, with room temperature perpendicular anisotropy and an enhanced surface scattering, were studied for the possible use in the extraordinary Hall effect (EHE) - based magnetic memory devices. Polarity of the EHE signal was found to change from negative in thick samples to positive in thin ones. Reversal of EHE sign was also observed in thick samples with aging. The effect is argued to be related to the dominance of surface scattering having the EHE polarity opposite to that of the bulk.

Extraordinary or anomalous Hall effect (EHE) was discovered about 130 years ago, however, discussion of its practical application only started recently when it was demonstrated [1] that a relatively small EHE coefficient of bulk ferromagnets can be boosted by stimulating proper scattering events. The addition of insulating impurities into bulk transition metals [2] or the increase of surface scattering by reducing the thickness of magnetic films below the mean free path [3] can enhance the effect by orders of magnitude. Thin films of ordinary transition metals (Ni, Co, Fe) with dominant surface scattering were shown to reach EHE resistivity of the order of 1 $\mu\Omega cm$ [1], whereas ultra-thin alloys of FePt [4] and CoFe/Pt multilayers [5] reach the values of about 5 $\mu\Omega cm$. Tuning the perpendicular anisotropy in CoFe/Pt multilayers, and thus reducing the magnetic saturation field below the typical demagnetizing value, was shown to provide the field sensitivity exceeding $10^3$ $\Omega$/T [5] which competes with the best semiconducting Hall sensors. Use of the EHE for magnetic memory devices with a direct data read-out is another potential application. Combination of an out-of-plane anisotropy with a strong remanent EHE signal has been demonstrated in thin Ni films [1, 6] at low temperatures. The goal of this work was the exploration of thin Co/Pd multilayers with room temperature perpendicular anisotropy in the thickness limit in which surface scattering can boost the magnitude of EHE, similar to the chemically homogeneous thin films. We shall demonstrate that the generation of room temperature magnetic memory units, and respectively, the magnetic random access memory based on the extraordinary Hall effect, is feasible. However, probably the most interesting and unexpected result we report on is the thickness and aging dependence of the EHE polarity in thin Co/Pd multilayers.

Several series of CoPd multilayers were fabricated by sequential electron-beam deposition of 0.2 nm thick Co and 0.9 nm thick Pd layers covered by 30 nm thick Ge on room temperature GaAs substrates. The number of repetitions (n) varied between 4 and 13. No post-deposition annealing was done. Films containing less than 4 bilayers were found non-conducting. Resistance and Hall effect were measured at room temperature up to 2 Tesla in both field polarities. The measured field dependent data was symmetrized and antisymmetrized to extract magnetoresistance and Hall resistivity respectively. The samples were repeatedly measured over a period of several months to track their time dependent evolution. EHE hysteresis loop of the thickest sample was compared with magnetization hysteresis loop extracted from the SQUID measurements. Like in other ferromagnetic films [1] and multilayers [7], the EHE signal was found to be an electrical replica of magnetization.

Fig. 1 presents the Hall resistivity of two multilayers with n equal 7 and 10 as a function of magnetic field applied normal to the film plane. Measurements were done shortly after the samples preparation. Both samples demonstrated square hysteresis loops with the remanent signal equal to the saturated EHE resistivity at high fields. Surprising is the change in polarity of the EHE component, which is positive in the n = 7 sample while negative for n = 10. Notably, the ordinary Hall effect coefficient, extracted from a linear slope at high fields above the magnetization saturation, is negative and equal to -0.01 μΩcm/T in all samples.

Zero field resistivity $\rho$ and the saturated EHE resistivity $\rho_{EHE}$ for a series of samples are shown in Fig. 2 as a function of their total thickness $t$. Resistivity increases with reduction of thickness due to an enhancement of the surface scattering [8]. Usually $\rho_{EHE}$ follows a similar thickness dependence as resistivity [1,3,6], however, in this case the polarity changes from negative in thick to positive in thin films. The same effect was also found at 77K. The coercive field $H_c$ is shown in the inset of Fig. 2 as a function of the repetition number n. No hysteresis was observed in the sample with n = 5, while in thicker samples $H_c$ grows linearly with n. This result is consistent with the previously published data [9, 10] and can be understood, either due to an increase of the uniaxial anisotropy, or the increment of the domain wall pinning centers created at Co-Pd interfaces.

Resistivity of all samples was found to increase after the deposition. The effect is common in thin films and can be related to a gradual oxidation and/or roughening of the surface [11]. Puzzling are the changes in the EHE resistivity. The effect is illustrated in Fig. 3 where we show three measurements of the same [Co/Pd]$_8$ sample taken 13, 37 and 61 days after the deposition. Polarity of the EHE signal gradually changes from negative immediately after the deposition to positive. Variation of EHE resistivity with time is positive in all samples regardless of the polarity of the signal shortly after the deposition. Thus, the absolute value of EHE resistivity of thin samples with originally positive EHE increased with time (see Fig.4), whereas in thick samples it decreased in an absolute value, then changed polarity from negative to positive and kept growing positive. Notably, the high field slope corresponding to the ordinary Hall effect component, squareness of the hysteresis loops, magnitudes of the coercive field (about 12mT), width of the signal reversal (about 2mT) remained constant (certain decrease of $H_c$ with time was observed in several samples depending on the deposition conditions).

There are several reports on the reversal of the EHE polarity, particularly in Co- based materials. The phenomenon was observed in CoPd alloys [12] and multilayers [13, 14] when concentration or a relative thickness ratio between the components was changed, and when a layer of FeMn antiferromagnet was added [15]. It was suggested [13] that the change of the position of Fermi level and, therefore shift of the majority carriers from electrons to holes, is the reason for the change of EHE polarity. This explanation is not supported by the ordinary Hall effect data that show no change in polarity and, therefore in the sort of charge carriers. A different approach [12, 15] is based on a traditional presentation of EHE as a superposition of two asymmetric scattering mechanisms: skew scattering and side jump in the form: $R_{EHE} = a\rho + b\rho^2$. Change of the composition can *a priori* result in a shift of a relative dominance of the linear and quadratic terms with coefficients *a, b* being of different polarities. In our case the ratio between Co and Pd is kept constant and only the total thickness is varied. Stability of the coercive field and squareness of hysteresis loops with aging also suggests that an interdiffusion at Co/Pd interfaces and therefore the composition changes are not significant.

Fig. 5 presents $\rho_{EHE}$ as a function of $\rho$ for a series of samples where data for each sample was taken over a period of about four months. All the data falls reasonably well on a single straight line, when thicker samples with lower resistivity change the polarity of $\rho_{EHE}$ from negative to positive. The result can be qualitatively understood by adapting an approach proposed in Ref. 16 where EHE resistivity is treated by using Matthiessen's rule. Similar to the longitudinal resistivity, the EHE resistivity can be presented as containing both bulk and surface scattering components. Assuming a simplified skew scattering mechanism with only linear dominant term, one gets the EHE resistivity in a form: $\rho_{EHE} = \rho_{EHE,b} + \rho_{EHE,s} = \alpha\rho_b + \beta\rho_s$, where $\rho_{EHE,b}$ and $\rho_{EHE,s}$ are the bulk and surface EHE resistivity components, and $\rho_b$ and $\rho_s$ are resistivity of the bulk and the surface respectively. Using the data presented in Figs.2 and 5 one can estimate the bulk and surface skew scattering coefficients as $\alpha \approx -10^{-3}$ and $\beta \approx 8\times10^{-3}$. The surface scattering coefficient $\beta$ is comparable with the values found in thin polycrystalline Ni films [3]: $2\times10^{-2}$ and $1.3\times10^{-2}$, (different values for differently fabricated series), epitaxial Fe films [17]: $5\times10^{-2}$ (although the authors of the latter paper analyzed their data differently), and granular Ni-SiO$_2$ mixtures [3]: $4\times10^{-3}$. The thickness and aging dependent variation of EHE resistivity can be attributed solely to the variation of the surface scattering. In thick films shortly after the deposition the surface scattering contribution is small, resistivity is independent of thickness and a negative polarity of EHE ($\alpha < 0$) is consistent with data reported for thick Co/Pd multilayers and alloys where Co volume content is below 25% [12, 13]. The surface scattering contribution to resistivity is dominant in thin films or becomes significant in thicker films with aging, probably because of gradual roughening of the surface and the enhancement of diffusive scattering. If the coefficient β is positive, a gradual increase of $\rho_s$ drives the polarity of $\rho_{EHE}$ from negative to positive.

To summarize, several parameters important for the room temperature applications of the EHE-based memory devices were tested in thin Co/Pd multilayers and found promising: EHE resistivity is of the order of 0.5 $\mu\Omega cm$, which is equivalent to a difference of the order of 1Ω between the up- and down- magnetized states; coercive field depends linearly on a number of bilayers and can be tailored in the field range of 10 - 1000 Gauss. Polarity of the EHE signal was found to depend on thickness and time elapsed since the deposition. We argue that the effect is determined by surface scattering in thin and aged samples. It would be beneficial to find a proper matching in polarity of spin-orbital scattering within and at the surface of the selected material, while in Co/Pd multilayers studied here these polarities are opposite.

This work was supported by the Israel Science Foundation grant No. 633/06. We thank V. Sheluchin for help in samples preparation.

# References.

1. A.Gerber, A.Milner, M.Karpovsky, B.Lemke, H.-U.Habermeier, J.Tuaillon-Combes, M.Negrier, O.Boisron, P.Melinon, and A.Perez, Jour. Magn. Magn. Matt. 242, 90 (2002);

2. A.B.Pakhomov, X.Yan, and B.Zhao, Appl. Phys. Lett. 67, 3497 (1995);

3. A.Gerber, A.Milner, LGoldsmith, M.Karpovsky, B.Lemke, H.-U.Habermeier, and A.Sulpice, Phys. Rev. B 65, 054426 (2002);

4. V.N. Matveev, V.I. Levashov, O.V. Kononenko, and A.N. Chaika, Russian Microelectronics 35, 392 (2006)

5. Y. Zhu and J.W. Cai, Appl. Phys. Lett. 90, 012104 (2007);

6. A. Gerber and O. Riss, J. Nanoelectron. Optoelectron. 3, 35 (2008);

7. C.L. Canedy, X.W. Li, and G. Xiao, Phys. Rev. B 62, 508 (2000);

8. K. Fuchs, Proc. Cambridge Phil. Soc. 34, 100 (1938);

9. T. Suzuki, Scripta Metallurgica et Materialia 33, 1609 (1995);

10. H.S. Lee, S.-B. Choe, S.-C. Shin, C.G. Kim, Jour. Mag. Mag. Mat. 239, 343 (2002);

11. S. Nakagawa, and H. Yoshikawa, Jour.Magn.Magn.Mat. 287, 193 (2005);

12. S.U. Jen, B.L. Chao, and C.C. Liu, J. Appl. Phys. 76, 5782 (1994);

13. S. Kim, S.R. Lee, and J.D. Chung, J. Appl. Phys. 73, 6344 (1993);

14. Y. Aoki, K. Honda, H. Sato, Y. Kobayashi, S. Hashimoto, T. Yokoyama, and T. Hanyu, Jour. Magn. Magn. Mat. 162 , 1 (1996);

15. C. Christides and Th. Speliotis, J. Appl. Phys. 97, 013901 (2005);

16. A.Gerber, A.Milner, A.Finkler, M.Karpovski, L.Goldsmith, J.Tuaillon-Combes, O.Boisron, P.Mélinon, and A.Perez., Phys. Rev. B 69, 224403 (2004);

17. Y.Tian, L.Ye, and X.Jin, Phys. Rev. Lett. 103, 087206 (2009).

**Figure captions.**

Fig.1. Hall resistivity as a function of magnetic field measured in two $(Co_{0.2\,nm}/Pd_{0.9\,nm})_n$ multilayers with the repetition number n = 7 (open circles) and n = 10 (solid circles) shortly after the deposition.

Fig.2. Resistivity $\rho$ (open circles) and EHE resistivity $\rho_{EHE}$ (solid circles) of a series of Co/Pd multilayers as a function of the total thickness. Inset: coercive field of the respective samples as a function of the repetition number.

Fig.3. Hall resistivity of the $[Co/Pd]_8$ sample (series 2) 13, 37 and 61 day after its deposition.

Fig.4. Resistivity $\rho$ (open circles) and EHE resistivity $\rho_{EHE}$ (solid circles) of the $[Co/Pd]_7$ sample as a function of time elapsed since its deposition.

Fig.5. EHE resistivity as a function of resistivity of several samples (series 1) with the repetition numbers n = 7, 8, 9, 10 and 12 measured 3, 15, 54 and 121 days after the deposition. Dashed line is the guide for eyes.

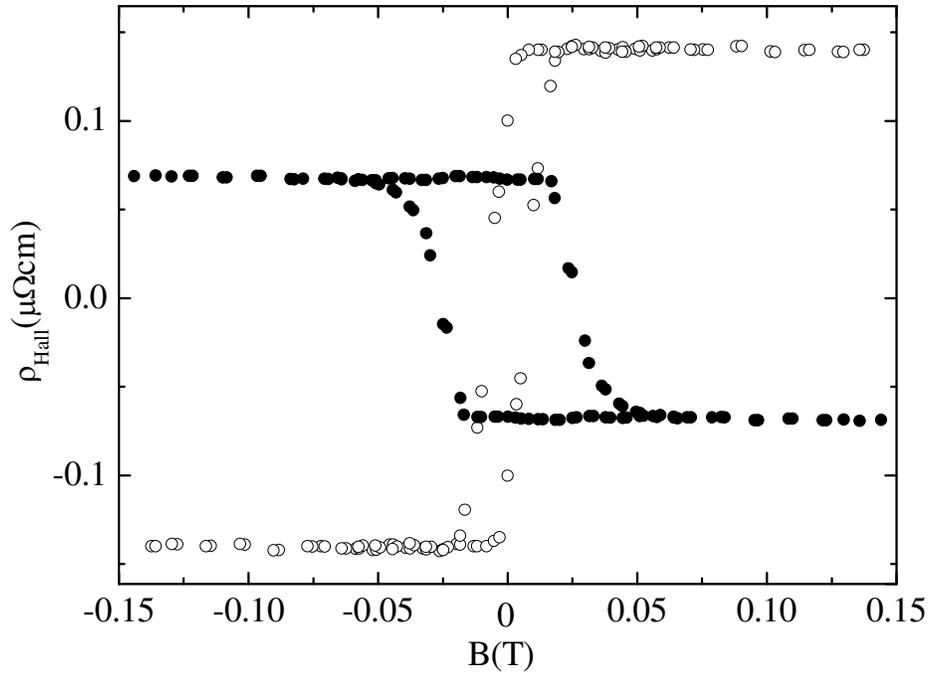

Fig.1

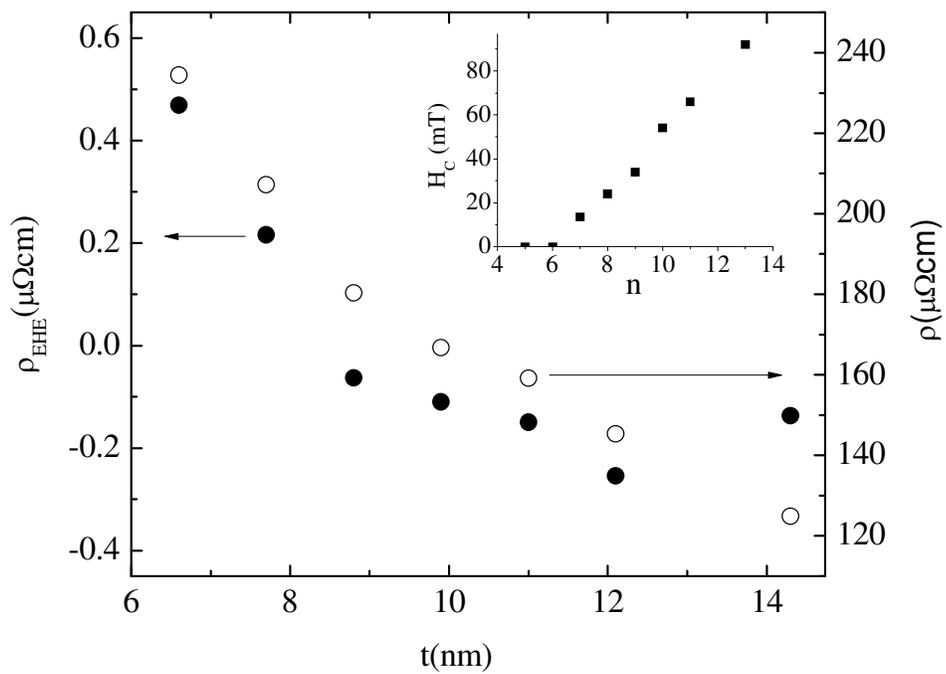

Fig.2

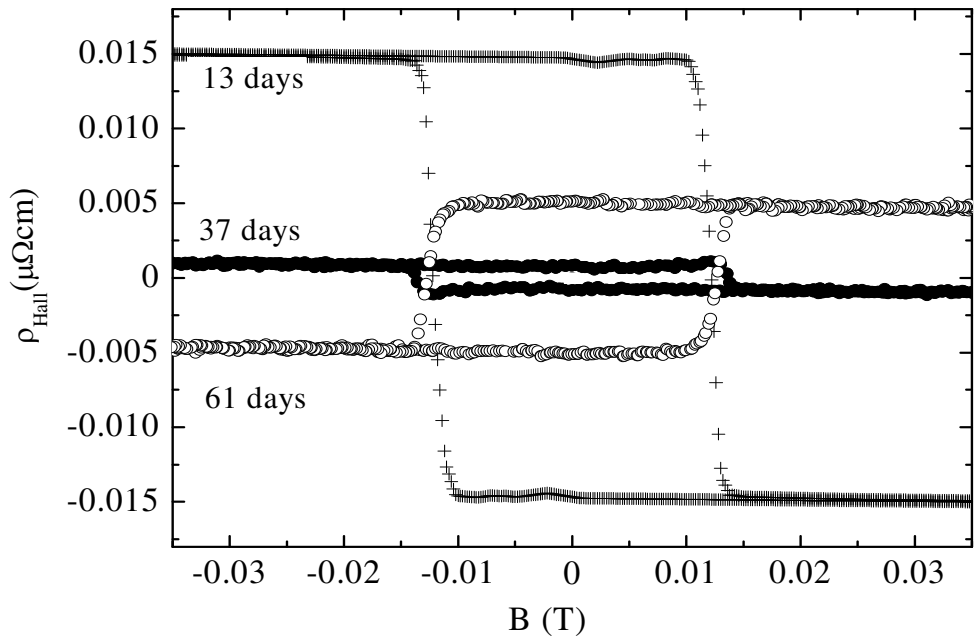

Fig.3

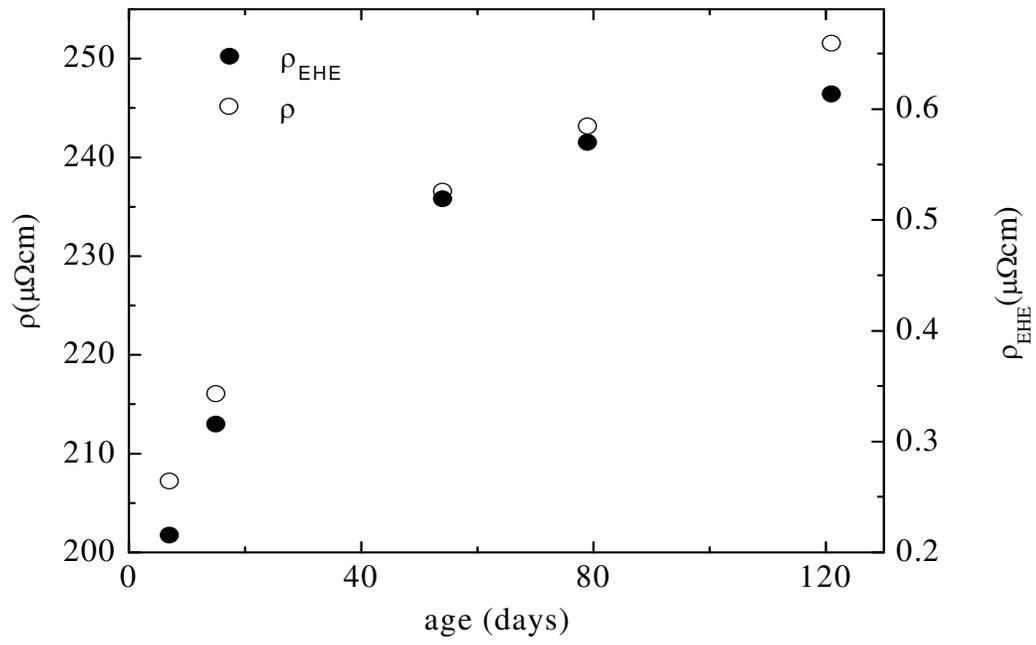

Fig.4

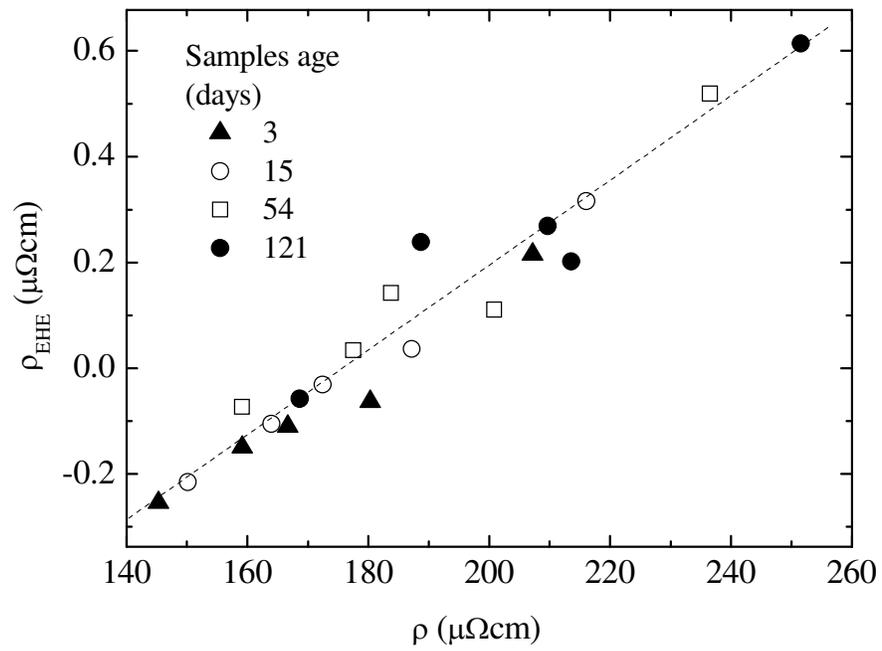

Fig.5